\documentclass[twoside]{IEEEtran}
\PassOptionsToPackage{ruled, linesnumbered}{algorithm2e}
\usepackage[latin9]{inputenc}
\usepackage{color}
\usepackage{mathrsfs}
\usepackage{algorithm2e}
\usepackage{amsmath}
\usepackage{amssymb}
\usepackage{graphicx}

\makeatletter

\usepackage{empheq}

\allowdisplaybreaks
\usepackage{pgfplots}
\usepackage{pgfplotstable}
\usepackage{tikz}
\usetikzlibrary{calc}
\usepackage{cite}
\usepackage{xcolor}
\definecolor{myblue}{rgb}{0.0, 0.5, 1.0}
\definecolor{myred}{rgb}{1.0, 0.13, 0.32}
\definecolor{mygreen}{rgb}{0.31, 0.78, 0.47}

\newcommand{\herm}{^{\mathsf{H}}}
\newcommand{\trans}{^{\mathsf{T}}}

\DeclareMathOperator{\minimize}{minimize}
\DeclareMathOperator{\diag}{diag}
\DeclareMathOperator{\maximize}{maximize}
\DeclareMathOperator{\st}{subject~to}

\makeatother

\begin{document}
\title{\huge{A Novel SCA-Based Method for Beamforming Optimization in IRS/RIS-Assisted
MU-MISO Downlink}}
\author{Vaibhav Kumar, \textit{Member, IEEE}, Rui Zhang, \textit{Fellow, IEEE},\\
Marco Di Renzo, \textit{Fellow,} \textit{IEEE}, and Le-Nam Tran, \textit{Senior
Member, IEEE}\thanks{This work was supported by a Grant from Science Foundation Ireland
under Grant 17/CDA/4786. The work of Rui Zhang was supported in part
by Ministry of Education, Singapore under Award T2EP50120-0024, Advanced
Research and Technology Innovation Centre (ARTIC) of National University
of Singapore under Research Grant R-261-518-005-720, and The Guangdong
Provincial Key Laboratory of Big Data Computing. The work of Marco
Di Renzo was supported in part by the European Commission through
the H2020 ARIADNE project under grant agreement number 871464 and
through the H2020 RISE-6G project under grant agreement number 101017011.\protect \\
Vaibhav Kumar, and Le-Nam Tran are with School of Electrical and Electronic
Engineering, University College Dublin, D04 V1W8 Dublin, Ireland (e-mail:
vaibhav.kumar@ieee.org; nam.tran@ucd.ie).\protect \\
R. Zhang is with The Chinese University of Hong Kong, Shenzhen, and
Shenzhen Research Institute of Big Data, Shenzhen, China 518172 (e-mail:
rzhang@cuhk.edu.cn). He is also with the Department of Electrical
and Computer Engineering, National University of Singapore, Singapore
117583 (e-mail: elezhang@nus.edu.sg).\protect \\
Marco Di Renzo is with Universit\'e Paris-Saclay, CNRS, CentraleSup\'elec,
Laboratoire des Signaux et Syst\'emes, 3 Rue Joliot-Curie, 91192
Gif-sur-Yvette, France (e-mail: marco.di-renzo@universite-paris-saclay.fr).}}
\maketitle
\begin{abstract}
In this letter, we consider the fundamental problem of jointly designing
the transmit beamformers and the phase-shifts of the intelligent reflecting
surface (IRS) / reconfigurable intelligent surface (RIS) to minimize
the transmit power, subject to quality-of-service constraints at individual
users in an IRS-assisted multiuser multiple-input single-output downlink
communication system. In particular, we propose a new \textit{successive
convex approximation based} \textit{second-order cone programming}
approach in which \textit{all the optimization variables are simultaneously
updated in each iteration}. Our proposed scheme achieves superior
performance compared to state-of-the-art benchmark solutions. In addition,
the complexity of the proposed scheme is $O(N_{\mathrm{s}}^{3.5})$,
while that of state-of-the-art benchmark schemes is $O(N_{\mathrm{s}}^{7})$,
where $N_{\mathrm{s}}$ denotes the number of reflecting elements
at the IRS.
\end{abstract}


\section{Introduction}

With the recent advancement in metamaterials, intelligent reflecting
surfaces (IRSs) or reconfigurable intelligent surfaces (RISs) are
being envisioned as one of the key enabling technologies for the next-generation
wireless communication systems~\cite{16-Nature,20-JSAC-Marco,21-ComMag-IRSfor6G,21-TCOM-ZhangTutorial,22-Zhang-Proc}.
The problem of transmit power minimization (PowerMin) for IRS-assisted
systems is of fundamental interest to reduce the system power consumption~\cite{20-ComMag-Zhang}.
In this context, the authors of~\cite{21-TCOM-SISO-BC,21-TCOM-SIMO-MAC,21-WCNC-IoT}
considered the PowerMin problem in IRS-assisted multiuser single-input
single-output (MU-SISO) downlink and multiuser single-input multiple-output
(MU-SIMO) uplink systems. Compared to traditional systems, the PowerMin
problem is more challenging to solve in the presence of IRSs due to
the coupling between the transmit beamformer(s) and the phase shifts
of the IRS. In particular, the PowerMin problem in IRS-assisted multiuser
multiple-input single-output (MU-MISO) downlink systems for different
scenarios, including multicasting~\cite{21-TCOM-Renzo}, broadcasting~\cite{19-TWC-Zhang-TPM},
symbol-level precoding~\cite{21-TWC-SymbolLevelPrecoding}, millimeter-wave
systems~\cite{21-WCNC-MISOBC}, and simultaneous wireless information
and power transfer (SWIPT) system~\cite{20-JSAC-Zhang-SWIPT}, were
recently addressed in the literature.

In the existing literature~\cite{19-TWC-Zhang-TPM,20-JSAC-Zhang-SWIPT,21-WCNC-IoT,21-WCNC-MISOBC,21-TWC-SymbolLevelPrecoding,21-TCOM-Renzo},
the prevailing method for solving the PowerMin problem is to alternately
optimize the transmit beamformer and the IRS phase-shifts, with one
of them kept fixed. Although alternating optimization (AO) greatly
simplifies the optimization problem, it may not yield a high-performance
solution due to the intricate coupling among the design variables.
From the point of view of the computational complexity, another limitation
of existing solutions to the PowerMin problem in IRS-assisted MU-MISO
systems is the extensive use of semidefinite relaxation (SDR)~\cite[(P7)]{21-TCOM-Renzo},~\cite[(P4')]{19-TWC-Zhang-TPM}
which incurs very high computational complexity, and is thus not suitable
for large-scale systems.

Our main purpose in this letter is to derive a more numerically efficient
solution to the PowerMin problem in IRS-assisted MU-MISO systems.
Specifically, we propose a provably-convergent successive convex approximation
(SCA) -based second-order cone programming (SOCP) approach, where
\textit{the transmit beamformers and IRS phase-shifts are updated
simultaneously in each iteration}. The proposed method is numerically
shown to achieve superior performance compared to a well-known \textcolor{black}{benchmark}
presented in~\cite{19-TWC-Zhang-TPM}, especially when the quality-of-service
(QoS) constraints are more demanding. We also show that the complexity
of the proposed scheme is significantly lower than the benchmark scheme
in~\cite{19-TWC-Zhang-TPM}, and therefore needs a considerably less
run time to compute a solution. While the PowerMin problem is specifically
studied in this letter, we also discuss the applicability of the proposed
method to other design problems in IRS-assisted communication systems.

\paragraph*{Notations}

Bold uppercase and lowercase letters denote matrices and vectors,
respectively. The (ordinary) transpose, conjugate transpose, Euclidean
norm, real component and imaginary component for a matrix $\mathbf{X}$
are denoted by $\mathbf{X}\trans$, $\mathbf{X}\herm$, $\Vert\mathbf{X}\Vert$,
$\Re\{\mathbf{X}\}$ and $\Im\{\mathbf{X}\}$, respectively. For a
complex number $x$, its absolute value is denoted by $|x|$. The
vector space of all complex-valued matrices of size $M\times N$ is
denoted by $\mathbb{C}^{M\times N}$. $\diag(\mathbf{x})$ denotes
the diagonal matrix whose main diagonal comprises the elements of
$\mathbf{x}$.

\section{Problem Formulation and State-of-The-Art \textcolor{black}{Solution}}

\begin{figure}[t]
\begin{centering}
\includegraphics[width=0.65\columnwidth]{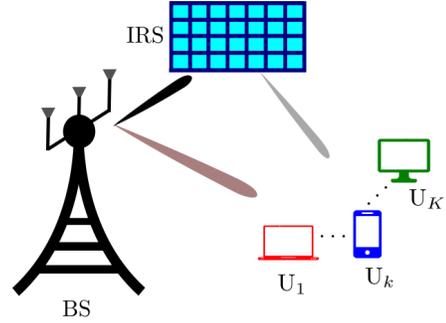}
\par\end{centering}
\caption{System model for IRS-assisted MU-MISO downlink.}
\label{fig:SysMod}
\end{figure}

Consider the IRS-assisted MU-MISO downlink system shown in~Fig.~\ref{fig:SysMod},
consisting of an $N_{\mathrm{t}}$-antenna base station (BS), one
IRS having $N_{\mathrm{s}}$ nearly-passive reflecting elements, and
$K$ single-antenna users (denoted by $\mathrm{U}_{k},k\in\mathcal{K}\triangleq\{1,2,\ldots,K\}$).
The BS-$\mathrm{U}_{k}$, BS-IRS and IRS-$\mathrm{U}_{k}$ links are
denoted by $\mathbf{h}_{\mathrm{t}k}\in\mathbb{C}^{1\times N_{\mathrm{t}}}$,
$\mathbf{H}_{\mathrm{ts}}\in\mathbb{C}^{N_{\mathrm{s}}\times N_{\mathrm{t}}}$
and $\mathbf{h}_{\mathrm{s}k}\in\mathbb{C}^{1\times N_{\mathrm{s}}}$,
respectively. The IRS reflection-coefficient vector is denoted by
$\boldsymbol{\phi}\triangleq[\phi_{1},\phi_{2},\ldots\phi_{N_{\mathrm{s}}}]\trans\in\mathbb{C}^{N_{\mathrm{s}}\times1}$,
such that $\phi_{n_{\mathrm{s}}}=\exp(-j\theta_{n_{\mathrm{s}}}),\forall n_{\mathrm{s}}\in\mathscr{N}_{\mathrm{s}}\triangleq\{1,2,\ldots,N_{\mathrm{s}}\}$
and $\theta_{n_{\mathrm{s}}}\in[0,2\pi)$ is the phase-shift of the
$n_{\mathrm{s}}$-th reflecting element. Let $s_{k}$ denote the information-bearing
symbol transmitted by the BS and intended for $\mathrm{U}_{k}$, and
$\mathbf{w}_{k}\in\mathbb{C}^{N_{\mathrm{t}}\times1}$ denote the
corresponding transmit beamforming/precoding vector. The signal received
at $\mathrm{U}_{k}$ is thus given by
\begin{equation}
y_{k}=(\mathbf{h}_{\mathrm{t}k}+\mathbf{h}_{\mathrm{s}k}\boldsymbol{\Phi}\mathbf{H}_{\mathrm{ts}})\sum\nolimits _{l\in\mathcal{K}}\mathbf{w}_{l}s_{l}+\omega_{k},\label{eq:RxSignal}
\end{equation}
where $\boldsymbol{\Phi}\triangleq\diag(\boldsymbol{\phi})$, and
$\omega_{k}\sim\mathcal{CN}\big(0,\sigma_{k}^{2}\big)$ denotes the
circularly-symmetric complex additive white Gaussian noise at $\mathrm{U}_{k}$,
with zero mean and variance of $\sigma_{k}^{2}$. Without loss of
generality, hereafter we assume that $\sigma_{k}=\sigma,\forall k\in\mathcal{K}$.
Also, to avoid any potential numerical issue when dealing with extremely
small quantities, in the rest of this letter, we make the normalization
$\mathbf{H}_{\mathrm{ts}}\leftarrow\mathbf{H}_{\mathrm{ts}}/\sigma$
and $\mathbf{h}_{\mathrm{t}k}\leftarrow\mathbf{h}_{\mathrm{t}k}/\sigma$.
Hence, the signal-to-interference-plus-noise ratio (SINR) at $\mathrm{U}_{k}$
is given by
\begin{equation}
\gamma_{k}=\frac{|\mathbf{g}_{k}\mathbf{w}_{k}|^{2}}{1+\sum\nolimits _{l\in\mathcal{K}\setminus\{k\}}|\mathbf{g}_{k}\mathbf{w}_{l}|^{2}},\label{eq:SINR-Def}
\end{equation}
where $\mathbf{g}_{k}\triangleq\mathbf{h}_{\mathrm{t}k}+\mathbf{h}_{\mathrm{s}k}\boldsymbol{\Phi}\mathbf{H}_{\mathrm{ts}},\forall k\in\mathcal{K}$.
We are interested in solving the fundamental problem of PowerMin by
jointly optimizing the transmit beamforming vectors ($\mathbf{w}\triangleq[\mathbf{w}_{1}\trans,\mathbf{w}_{2}\trans,\ldots,\mathbf{w}_{K}\trans]\trans$)
and the IRS phase-shift vector $\boldsymbol{\phi}$, such that a minimum
required SINR is always maintained at each of the users. Specifically,
the optimization problem can be formulated as\begin{subequations}\label{eq:OptProb-Orig} \begin{align}   
\underset{\mathbf{w},\boldsymbol{\phi}}{\minimize}\  & 
\Vert\mathbf{w}
\Vert^{2},\label{eq:OptObj-Orig}\\ \st\  & \gamma_{k}\geq\Gamma_{k},\quad\forall k\in\mathcal{K},\label{eq:SINRC-Orig}\\  & |\phi_{n_{\mathrm{s}}}|=1,\quad\forall n_{\mathrm{s}}\in\mathscr{N}_{\mathrm{s}},\label{eq:UMC-Orig}
\end{align} \end{subequations}where $\Gamma_{k}$ denotes the minimum required SINR that needs
to be met at $\mathrm{U}_{k}$ to provide a minimum acceptable QoS.
To characterize the full potential of IRS, in this letter, we assume
that perfect channel-state information (CSI) for all of the links
is available at the BS.\footnote{A similar assumption regarding the perfect CSI availability in different
IRS-assisted systems was considered in~\cite{21-TCOM-SIMO-MAC,21-WCNC-IoT,21-TCOM-Renzo,19-TWC-Zhang-TPM,21-TWC-SymbolLevelPrecoding,21-WCNC-MISOBC,20-JSAC-Zhang-SWIPT,21-Stefan-MIMOBC,Vaibhav_SpectrumSharing}.
The results in this letter serve as theoretical performance bounds
for the IRS-MU-MIMO system with imperfect CSI. An overview of the
available methods for acquiring the CSI and for designing IRS-assisted
systems for different levels of CSI availability is available in~\cite{22-JSTSP-Pan,22-COMST-Zhang}.}

Problem~\eqref{eq:OptProb-Orig} was first studied in~\cite{19-TWC-Zhang-TPM},
where an AO-based approach was proposed. More specifically, for fixed
$\boldsymbol{\phi}$, the transmit beamforming vector was updated
by solving the following problem
\begin{equation}
\underset{\mathbf{w}}{\minimize}\ \bigl\{\Vert\mathbf{w}\Vert^{2}\ \bigl|\ \eqref{eq:SINRC-Orig}\bigl\}.\label{eq:w-update}
\end{equation}
Then, for fixed $\mathbf{w}$, $\boldsymbol{\phi}$ was updated as
the solution to the following problem:
\begin{gather}
\underset{\boldsymbol{\phi},\{\alpha_{k}\}_{k\in\mathcal{K}}}{\maximize}\ \bigl\{\sum\nolimits _{k\in\mathcal{K}}\alpha_{k}\ \bigl|\ |\mathbf{g}_{k}\mathbf{w}_{k}|^{2}\geq\gamma_{k}(1+\sum_{l\in\mathcal{K}\setminus\{k\}}|\mathbf{g}_{k}\mathbf{w}_{l}|^{2})\nonumber \\
+\alpha_{k},\forall k\in\mathcal{K};\eqref{eq:UMC-Orig}\bigl\}.\label{eq:theta-update}
\end{gather}
While there exist several efficient methods for solving~\eqref{eq:w-update},
problem~\eqref{eq:theta-update} is a non-convex quadratically constrained
quadratic program, and thus is difficult to be solved optimally. The
authors of \cite{19-TWC-Zhang-TPM}\textbf{ }applied the SDR with
Gaussian randomization methods~\cite{06-TSP-ZhiQuan,18-Book-Bjorn}
to solve~\eqref{eq:theta-update}. There are two main drawbacks for
such an approach. First, the complexity of solving~\eqref{eq:theta-update}
using SDR increases very quickly with the number of reflecting elements
since the problem is lifted to the positive semidefinite domain by
defining $\mathbf{V}_{k}=\mathbf{w}_{k}\mathbf{w}_{k}\herm\in\mathbb{C}^{N_{\mathrm{s}}\times N_{\mathrm{s}}}$.
Second, extracting a rank-1 feasible solution from $\mathbf{V}_{k}$
requires a large number of randomization steps which significantly
adds to the overall complexity. More importantly, as~\eqref{eq:SINRC-Orig}
results in a high degree of coupling between $\boldsymbol{\phi}$
and $\mathbf{w}$, AO-based methods are usually not efficient since
a stationary point is not guaranteed to be obtained theoretically.

\section{\label{sec:Proposed-Solution}Proposed Solution}

It is now clear that the non-convexity of~\eqref{eq:SINRC-Orig}
is the main issue to address. In this section, we apply the SCA framework
to tackle this problem by proposing a series of convex approximations.
To this end, we recall the following inequality and equalities:\begin{subequations} \begin{align}   
\Vert\mathbf{x}\Vert^{2} & \ \geq2\Re\{\mathbf{y}\herm\mathbf{x}\}-\Vert\mathbf{y}\Vert^{2},\label{eq:Inequality-1}\\
\Re\{\mathbf{x}\herm\mathbf{y}\} & \ =\tfrac{1}{4}\big(\Vert\mathbf{x}+\mathbf{y}\Vert^{2}-\Vert\mathbf{x}-\mathbf{y}\Vert^{2}\big),\label{eq:EqualityReal-1}\\
\Im\{\mathbf{x}\herm\mathbf{y}\} & \ =\tfrac{1}{4}\big(\Vert\mathbf{x}-j\mathbf{y}\Vert^{2}-\Vert\mathbf{x}+j\mathbf{y}\Vert^{2}\big),\label{eq:EqualityImag-1}
\end{align} \end{subequations}which hold for two arbitrary complex-valued vectors $\mathbf{x}$
and $\mathbf{y}$. Note that~\eqref{eq:Inequality-1} is obtained
by linearizing $\Vert\mathbf{x}\Vert^{2}$ around $\mathbf{y}$ and
the equality in~\eqref{eq:Inequality-1} occurs for $\mathbf{x}=\mathbf{y}$, and \eqref{eq:EqualityReal-1} and \eqref{eq:EqualityImag-1}
are obtained by expanding the terms in the right hand side.

We deal with \eqref{eq:SINRC-Orig} using the notion of SCA \emph{where
both $\boldsymbol{\phi}$ and $\mathbf{w}$ are optimized in each
iteration}. First, ~\eqref{eq:SINRC-Orig} is equivalent to
\begin{subequations}
\label{eq:SINR:equiv}
\begin{align}
\frac{|\mathbf{g}_{k}\mathbf{w}_{k}|^{2}}{\Gamma_{k}} & \geq1+\sum\nolimits _{l\in\mathcal{K}\setminus\{k\}}\big(t_{kl}^{2}+\bar{t}_{kl}^{2}\big),\label{eq:SINRC-1}\\
t_{kl} & \geq|\Re\{\mathbf{g}_{k}\mathbf{w}_{l}\}|,\quad\forall l\in\mathcal{K}\setminus\{k\},\label{eq:tklDef}\\
\bar{t}_{kl} & \geq\ |\Im\{\mathbf{g}_{k}\mathbf{w}_{l}\}|,\quad\forall l\in\mathcal{K}\setminus\{k\},\label{eq:tBarklDef}
\end{align}
\end{subequations}
where $t_{kl}$ and $\bar{t}_{kl}$ are newly introduced slack variables.
It is straightforward to see that if \eqref{eq:SINRC-Orig} is feasible,
then so is \eqref{eq:SINR:equiv} and vice versa.

Since the right-hand side (RHS) of~\eqref{eq:SINRC-1} is convex,
we need to find a \textit{concave lower bound} for the term $|\mathbf{g}_{k}\mathbf{w}_{k}|^{2}$
in~\eqref{eq:SINRC-1}. Let $\boldsymbol{\phi}^{(n)}$ and $\mathbf{w}_{k}^{(n)}$
represent the value of $\boldsymbol{\phi}$ and $\mathbf{w}_{k}$
in the $n$-th iteration of the SCA process, respectively. Then we
have
\begin{align}
 & |\mathbf{g}_{k}\mathbf{w}_{k}|^{2}\overset{(\mathrm{a})}{\geq}\ 2\Re\big\{\big(a_{k}^{(n)}\big)\herm\mathbf{g}_{k}\mathbf{w}_{k}\big\}-\big|a_{k}^{(n)}\big|^{2}\nonumber \\
\overset{(\mathrm{b})}{\geq}\  & \tfrac{1}{2}\big\{\big\Vert\big(a_{k}^{(n)}\big)\mathbf{g}_{k}\herm+\mathbf{w}_{k}\big\Vert^{2}-\big\Vert\big(a_{k}^{(n)}\big)\mathbf{g}_{k}\herm-\mathbf{w}_{k}\big\Vert^{2}\big\}-\big|a_{k}^{(n)}\big|^{2}\nonumber \\
\overset{(\mathrm{c})}{\geq}\  & \Re\big\{\big(\mathbf{b}_{k}^{(n)}\big)\herm\big[\big(a_{k}^{(n)}\big)\mathbf{g}_{k}\herm+\mathbf{w}_{k}\big]\big\}-\tfrac{1}{2}\big\Vert\mathbf{b}_{k}^{(n)}\big\Vert^{2}\nonumber \\
 & \qquad\qquad-\tfrac{1}{2}\big\Vert\big(a_{k}^{(n)}\big)\mathbf{g}_{k}\herm-\mathbf{w}_{k}\big\Vert^{2}-\big|a_{k}^{(n)}\big|^{2}\nonumber \\
\triangleq\  & f_{k}\big(\mathbf{w}_{k},\boldsymbol{\phi};\mathbf{w}_{k}^{(n)},\boldsymbol{\phi}^{(n)}\big),\label{eq:fDef}
\end{align}
where $a_{k}^{(n)}\triangleq\mathbf{g}_{k}^{(n)}\mathbf{w}_{k}^{(n)}$,
$\mathbf{b}_{k}^{(n)}\triangleq a_{k}^{(n)}\big(\mathbf{g}_{k}^{(n)}\big)\herm+\mathbf{w}_{k}^{(n)}$,
and $\mathbf{g}_{k}^{(n)}\triangleq\mathbf{h}_{\mathrm{t}k}+\mathbf{h}_{\mathrm{s}k}\diag(\boldsymbol{\phi}^{(n)})\mathbf{H}_{\mathrm{ts}}$.
Specifically, $(\mathrm{a})$ and $(\mathrm{c})$ are due to~\eqref{eq:Inequality-1},
and $(\mathrm{b})$ is due to~\eqref{eq:EqualityReal-1}. It is easy
to check that $f_{k}\big(\mathbf{w}_{k},\boldsymbol{\phi};\mathbf{w}_{k}^{(n)},\boldsymbol{\phi}^{(n)}\big)$
is \emph{jointly concave} with respect to (w.r.t.) $\boldsymbol{\phi}$
and $\mathbf{w}$.

Using the fact that $u\geq|v|$ if and only if $u\geq v$ and $u\geq-v$,
and using~\eqref{eq:EqualityReal-1} we can equivalently rewrite
\eqref{eq:tklDef} as
\begin{subequations}
\label{eq:tkl-Inequal-1}
\begin{align}
t_{kl}\geq\  & \Re\{\mathbf{g}_{k}\mathbf{w}_{l}\}=\tfrac{1}{4}\big(\Vert\mathbf{g}_{k}\herm+\mathbf{w}_{l}\Vert^{2}-\Vert\mathbf{g}_{k}\herm-\mathbf{w}_{l}\Vert^{2}\big),\label{eq:tkl-Inequal-1-1}\\
t_{kl}\geq\  & -\Re\{\mathbf{g}_{k}\mathbf{w}_{l}\}=\tfrac{1}{4}\big(\Vert\mathbf{g}_{k}\herm-\mathbf{w}_{l}\Vert^{2}-\Vert\mathbf{g}_{k}\herm+\mathbf{w}_{l}\Vert^{2}\big).\label{eq:tkl-Inequal-1-2}
\end{align}
\end{subequations}
The non-convexity of \eqref{eq:tkl-Inequal-1-1} is due to the negative
quadratic term in the RHS. To convexify \eqref{eq:tkl-Inequal-1-1},
we apply \eqref{eq:Inequality-1} with $\mathbf{x}=\mathbf{g}_{k}\herm-\mathbf{w}_{l}$
and\textbf{ }$\mathbf{y}=\mathbf{g}_{k}^{(n)}-\big(\mathbf{w}_{l}^{(n)}\big)\herm$$-\Vert\mathbf{g}_{k}\herm-\mathbf{w}_{l}\Vert^{2}$,
which yields
\begin{align}
t_{kl}\geq\  & \tfrac{1}{4}\big[\big\Vert\mathbf{g}_{k}\herm+\mathbf{w}_{l}\big\Vert^{2}-2\Re\big\{\big(\mathbf{g}_{k}^{(n)}-\big(\mathbf{w}_{l}^{(n)}\big)\herm\big)\big(\mathbf{g}_{k}\herm-\mathbf{w}_{l}\big)\big\}\nonumber \\
 & \!\!\!\!\!+\big\Vert\big(\mathbf{g}_{k}^{(n)}\big)\herm\!-\!\mathbf{w}_{l}^{(n)}\big\Vert^{2}\big]\!\triangleq\!\mu_{kl}\big(\mathbf{w}_{l},\boldsymbol{\phi};\mathbf{w}_{l}^{(n)},\boldsymbol{\phi}^{(n)}\big).\label{eq:mukl}
\end{align}
Note that the RHS in~\eqref{eq:mukl} is \emph{jointly convex} w.r.t.
$\boldsymbol{\phi}$ and $\mathbf{w}_{l}$, resulting in the convexity
of $\mu_{kl}\big(\mathbf{w}_{l},\boldsymbol{\phi};\mathbf{w}_{l}^{(n)},\boldsymbol{\phi}^{(n)}\big)$.
Following a similar line of arguments, we can approximate~\eqref{eq:tkl-Inequal-1-2}
as
\begin{align}
t_{kl}\geq\  & \tfrac{1}{4}\big[\big\Vert\mathbf{g}_{k}\herm-\mathbf{w}_{l}\big\Vert^{2}-2\Re\big\{\big(\mathbf{g}_{k}^{(n)}+\big(\mathbf{w}_{l}^{(n)}\big)\herm\big)\big(\mathbf{g}_{k}\herm+\mathbf{w}_{l}\big)\big\}\nonumber \\
 & \!\!\!\!\!+\big\Vert\big(\mathbf{g}_{k}^{(n)}\big)\herm\!+\!\mathbf{w}_{l}^{(n)}\big\Vert^{2}\big]\!\triangleq\!\hat{\mu}_{kl}\big(\mathbf{w}_{l},\boldsymbol{\phi};\mathbf{w}_{l}^{(n)},\boldsymbol{\phi}^{(n)}\big).\label{eq:muBarkl}
\end{align}
Similarly,~\eqref{eq:tBarklDef} leads to the following two inequalities:
\begin{subequations}
\label{eq:tBarkl-Inequal-1}
\begin{align}
\bar{t}_{kl}\geq\  & \Im\{\mathbf{g}_{k}\mathbf{w}_{l}\}=\tfrac{1}{4}\big(\big\Vert\mathbf{g}_{k}\herm-j\mathbf{w}_{l}\big\Vert^{2}-\big\Vert\mathbf{g}_{k}\herm+j\mathbf{w}_{l}\big\Vert^{2}\big),\label{eq:tBarkl-Inequal-1-1}\\
\bar{t}_{kl}\geq\  & -\Im\{\mathbf{g}_{k}\mathbf{w}_{l}\}\!=\!\tfrac{1}{4}\big(\big\Vert\mathbf{g}_{k}\herm\!+\!j\mathbf{w}_{l}\big\Vert^{2}\!-\!\big\Vert\mathbf{g}_{k}\herm\!-\!j\mathbf{w}_{l}\big\Vert^{2}\big).\label{eq:tBarkl-Inequal-1-2}
\end{align}
\end{subequations}
Using the same line of thought of~\eqref{eq:mukl}, we can obtain
a lower bound for~\eqref{eq:tBarkl-Inequal-1} as follows:
\begin{align}
\bar{t}_{kl}\geq\  & \tfrac{1}{4}\big[\big\Vert\mathbf{g}_{k}\herm\!-\!j\mathbf{w}_{l}\big\Vert^{2}\!-\!2\Re\big\{\big(\mathbf{g}_{k}^{(n)}\!-\!j\big(\mathbf{w}_{l}^{(n)}\big)\herm\big)\big(\mathbf{g}_{k}\herm\!+\!j\mathbf{w}_{l}\big)\big\}\nonumber \\
 & \!\!\!\!\!+\big\Vert\big(\mathbf{g}_{k}^{(n)}\big)\herm\!+\!j\mathbf{w}_{l}^{(n)}\big\Vert^{2}\big]\!\triangleq\!\nu_{kl}\big(\mathbf{w}_{l},\boldsymbol{\phi};\mathbf{w}_{l}^{(n)},\boldsymbol{\phi}^{(n)}\big),\!\!\label{eq:nukl}\\
\bar{t}_{kl}\geq\  & \tfrac{1}{4}\big[\big\Vert\mathbf{g}_{k}\herm\!+\!j\mathbf{w}_{l}\big\Vert^{2}\!-\!2\Re\big\{\big(\mathbf{g}_{k}^{(n)}\!+\!j\big(\mathbf{w}_{l}^{(n)}\big)\herm\big)\big(\mathbf{g}_{k}\herm\!-\!j\mathbf{w}_{l}\big)\big\}\nonumber \\
 & \!\!\!\!\!+\big\Vert\big(\mathbf{g}_{k}^{(n)}\big)\herm\!-\!j\mathbf{w}_{l}^{(n)}\big\Vert^{2}\big]\!\triangleq\!\hat{\nu}_{kl}\big(\mathbf{w}_{l},\boldsymbol{\phi};\mathbf{w}_{l}^{(n)},\boldsymbol{\phi}^{(n)}\big).\!\!\label{eq:nuBarkl}
\end{align}
\begin{algorithm}[t]
\caption{Proposed SCA-based Method to Solve~\eqref{eq:OptProb-Orig}.}

\label{algo}

\KwIn{$\mathbf{w}^{(0)}$, $\boldsymbol{\phi}^{(0)}$, $\xi>0$}

$n\leftarrow0$\;

\Repeat{convergence }{

Solve ~\eqref{eq:Pn1} and denote the solution as $\mathbf{w}^{\star}$,
$\boldsymbol{\phi}^{\star}$\;

Update: $\mathbf{w}^{(n+1)}\leftarrow\mathbf{w}_{k}^{\star}$, $\boldsymbol{\phi}^{(n+1)}\leftarrow\boldsymbol{\phi}^{\star}$\;

$n\leftarrow n+1$\;

}

\KwOut{$\mathbf{w}^{\star}$, $\boldsymbol{\phi}^{\star}$}
\end{algorithm}
We are left with the non-convexity of~\eqref{eq:UMC-Orig}. To avoid
increasing the problem size, we tackle~\eqref{eq:UMC-Orig} by first
relaxing it to be an inequality (and thus convex) constraint, and
then forcing the inequality to be an equality by adding a regularized
term to the objective function in~\eqref{eq:OptObj-Orig}, which
results in the following optimization problem:
\begin{subequations}
\label{eq:OptProb-Penalized}
\begin{align}
\underset{\mathbf{w},\boldsymbol{\phi}}{\minimize} & \ \Vert\mathbf{w}\Vert^{2}-\xi\Vert\boldsymbol{\phi}\Vert^{2}\label{eq:OptObj-Penalized}\\
\st & \ f_{k}\big(\mathbf{w}_{k},\boldsymbol{\phi};\mathbf{w}_{k}^{(n)},\boldsymbol{\phi}^{(n)}\big)\nonumber \\
 & \quad\geq1+\sum\nolimits _{l\in\mathcal{K}\setminus\{k\}}\big(t_{kl}^{2}+\bar{t}_{kl}^{2}\big),\forall k\in\mathcal{K},\label{eq:SINR:final}\\
 & \ \eqref{eq:mukl},\eqref{eq:muBarkl},\eqref{eq:nukl},\eqref{eq:nuBarkl},\forall k,l\in\mathcal{K},k\neq l,\label{eq:groupedInequality}\\
 & \ |\phi_{n_{\mathrm{s}}}|\leq1,\quad\forall n_{\mathrm{s}}\in\mathscr{N}_{\mathrm{s}},\label{eq:UMC-Relaxed}
\end{align}
\end{subequations}
where $\xi>0$ is the regularization parameter. It can be shown that
for sufficiently large $\xi$,~\eqref{eq:UMC-Relaxed} is binding
at the optimality when convergence is achieved for the problem in~\eqref{eq:OptProb-Penalized}.
Note that due to the concavity of the term $-\xi\Vert\boldsymbol{\phi}\Vert^{2}$,
the objective in~\eqref{eq:OptProb-Penalized} is still non-convex.
However, we can convexify~\eqref{eq:OptObj-Penalized} using~\eqref{eq:Inequality-1}.
In summary, the approximate convex subproblem for~\eqref{eq:OptProb-Orig}
in the $(n\!+\!1)$-th iteration of the SCA process is given by

\begin{equation}
\begin{aligned}
\underset{\mathbf{w},\boldsymbol{\phi},\mathbf{t},\bar{\mathbf{t}}}{\minimize}  & \  \Vert\mathbf{w}\Vert^{2}
-\!\xi\big[2\Re\big\{\big(\boldsymbol{\phi}^{(n)}\big)\herm\boldsymbol{\phi}\big\}\!-\!\Vert\boldsymbol{\phi}^{(n)}\Vert^{2}\big] \\
\st & \ \eqref{eq:SINR:final},\eqref{eq:groupedInequality},\eqref{eq:UMC-Relaxed},
\end{aligned}\tag{$\ensuremath{\mathcal{P}_{n+1}}$}\label{eq:Pn1}
\end{equation}where $\mathbf{t}=\{t_{kl}\}_{k,l\in\mathcal{K},k\neq l}$ and $\bar{\mathbf{t}}=\{\bar{t}_{kl}\}_{k,l\in\mathcal{K},k\neq l}$.
Finally, the proposed SCA-based method for solving~\eqref{eq:OptProb-Orig}
is outlined in \textbf{Algorithm~\ref{algo}}. To obtain the initial
points $(\mathbf{w}^{(0)},\boldsymbol{\phi}^{(0)})$ we randomly generate
$\boldsymbol{\phi}^{(0)}$ and then find $\mathbf{w}^{(0)}$ by solving~\eqref{eq:w-update}.
We also note that~\eqref{eq:Pn1} is an SOCP problem, and can be
solved efficiently using off-the-shelf solvers such as MOSEK~\cite{mosek}.

\subsection{Convergence Analysis}

Let $f(\mathbf{w},\boldsymbol{\phi})=\Vert\mathbf{w}\Vert^{2}-\xi\Vert\boldsymbol{\phi}\Vert^{2}$.
For a given $\xi$, since the optimal solution of $(\mathcal{P}_{n})$
is also a feasible solution to $(\mathcal{P}_{n+1})$, Algorithm \ref{algo}
yields a non-increasing objective sequence, i.e. $f(\mathbf{w}^{(n)},\boldsymbol{\phi}^{(n)})\geq f(\mathbf{w}^{(n+1)},\boldsymbol{\phi}^{(n+1)})$.
Due to \eqref{eq:UMC-Relaxed}, the objective sequence is bounded
from below, i.e. $f(\mathbf{w},\boldsymbol{\phi})\geq-\xi N_{\mathrm{s}}$
and thus $f(\mathbf{w}^{(n)},\boldsymbol{\phi}^{(n)})$ is convergent.
We can prove that the sequence of iterates is also convergent as follows.
Let $\mathcal{X}^{(0)}=\{\mathbf{w},\boldsymbol{\phi}|f(\mathbf{w},\boldsymbol{\phi})\leq f(\mathbf{w}^{(0)},\boldsymbol{\phi}^{(0)})$
be the level set associated with the initial point. Note that $\mathcal{X}^{(0)}$
is closed. We further note that $f(\mathbf{w},\boldsymbol{\phi})\leq f(\mathbf{w}^{(0)},\boldsymbol{\phi}^{(0)})$
implies $\Vert\mathbf{w}\Vert^{2}\leq f(\mathbf{w}^{(0)},\boldsymbol{\phi}^{(0)})+\xi\Vert\boldsymbol{\phi}\Vert^{2}\leq f(\mathbf{w}^{(0)},\boldsymbol{\phi}^{(0)})+\xi N_{\mathrm{s}}$,
meaning $\mathcal{X}^{(0)}$ is bounded and thus compact. Since the
sequence $\{(\mathbf{w}^{(n)},\boldsymbol{\phi}^{(n)})\}$ must lie
in the compact set $\mathcal{X}^{(0)}$, it has a limit point. We
can prove that this limit point is a stationary point of the following
problem 
\[
\begin{aligned}\underset{\mathbf{w},\boldsymbol{\phi}}{\minimize}\ \big\{ & \Vert\mathbf{w}\Vert^{2}-\xi\Vert\boldsymbol{\phi}\Vert^{2}\big|\eqref{eq:SINRC-Orig},\eqref{eq:UMC-Relaxed}\big\}.\end{aligned}
\]
If \eqref{eq:UMC-Relaxed} occurs with equality which holds for large
$\xi$, then the limit point is also a stationary solution to~\eqref{eq:OptProb-Orig}.
Interestingly, our numerical results show that this is always the
case even for a small $\xi$. We remark that in the case~\eqref{eq:UMC-Relaxed}
is not satisfied when Algorithm~\ref{algo} converges, we can simply
scale $\boldsymbol{\phi}$ to satisfy~\eqref{eq:UMC-Orig} and solve~\eqref{eq:w-update}
to find the final solution.

\subsection{\label{subsec:Complexity-Analysis}Complexity Analysis}

Note that in~\eqref{eq:Pn1}, we introduce a slack variable in the
objective to replace the term $\Vert\mathbf{w}\Vert^{2}$. It is clear
from~\eqref{eq:Pn1} that the size (in the real-domain) of the optimization
variables is $2(KN_{\mathrm{t}}+N_{\mathrm{s}}+K(K-1))+1$, and the
total number of SOC constraints is $K+4K(K-1)+N_{\mathrm{s}}+1$.
Therefore, following~\cite[Sec.~6.6.2]{ModernOptLectures}, the overall
per-iteration complexity of the proposed SOCP-based algorithm is given
by 
\begin{align}
 & \!\!\!\mathcal{C}_{\mathrm{SOCP}}=\mathcal{O}\big[2\big(4K^{2}+N_{\mathrm{s}}\big)^{0.5}\big(K^{2}+KN_{\mathrm{t}}+N_{\mathrm{s}}\big)\big(4K^{5}\nonumber \\
 & \!\!\!\!\!+16K^{3}N_{\mathrm{t}}+8K^{2}N_{\mathrm{s}}+20K^{2}N_{\mathrm{t}}^{2}+8KN_{\mathrm{t}}N_{\mathrm{s}}+4N_{\mathrm{s}}^{2})\big].\label{eq:ComplexitySOCP_Exact}
\end{align}
In practice, an IRS usually has a large number of reflecting elements,
i.e., $N_{\mathrm{s}}\gg\max\{N_{\mathrm{t}},K\}$~\cite{20-IRS_Prototype,21-RIS-Prototype}.
Therefore, \textcolor{black}{using~\eqref{eq:ComplexitySOCP_Exact},
the }approximate \textcolor{black}{per-iteration complexity of the
proposed SOCP-based algorithm is given by $\mathcal{C}_{\mathrm{SOCP}}\approx\text{\ensuremath{\mathcal{O}\big(N_{\mathrm{s}}^{0.5}\times N_{\mathrm{s}}\times N_{\mathrm{s}}^{2}\big)}}=\mathcal{O}\big(N_{\mathrm{s}}^{3.5}\big)$.
On the other hand, since the complexity of~\cite[Algorithm~1]{19-TWC-Zhang-TPM}
is dominated by that of solving }\eqref{eq:theta-update} using SDP,
the per-iteration complexity of the SDR-based algorithm in~\cite[Algorithm~1]{19-TWC-Zhang-TPM}
is given by $\mathcal{C}_{\mathrm{SDR}}\approx\mathcal{O}\big(N_{\mathrm{s}}^{7}\big)$
\cite[Sec.~6.6.2]{ModernOptLectures}, which is significantly higher
than that of our proposed algorithm. This will be numerically shown
in the next section.

\begin{figure}[t]
\begin{centering}
\includegraphics[width=0.7\columnwidth]{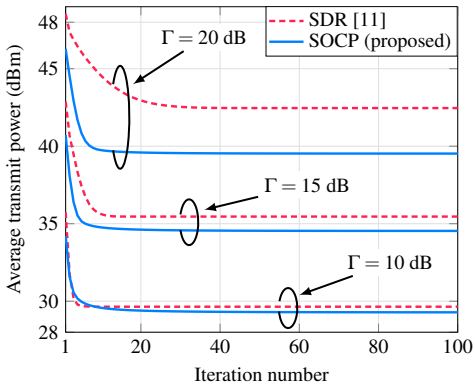}\vspace{-0.2cm}
\par\end{centering}
\caption{Convergence results for $N_{\mathrm{t}}=K=4$ and $N_{\mathrm{s}}=100$.}
\label{fig:convergence}
\end{figure}

\begin{figure}[t]
\begin{centering}
\includegraphics[width=0.7\columnwidth]{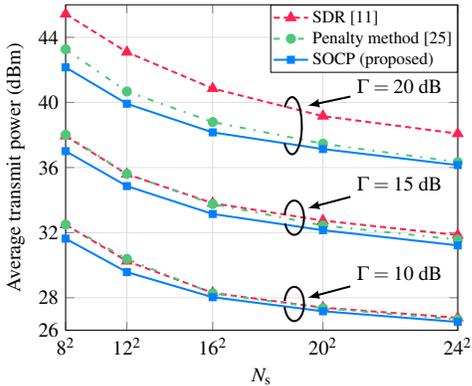}\vspace{-0.2cm}
\par\end{centering}
\caption{Average transmit power versus $N_{\mathrm{s}}$ \textcolor{black}{for
$N_{\mathrm{t}}=K=6$}.}
\label{fig:power_vs_Ns}
\end{figure}

\begin{figure}[t]
\centering{}\includegraphics[width=0.7\columnwidth]{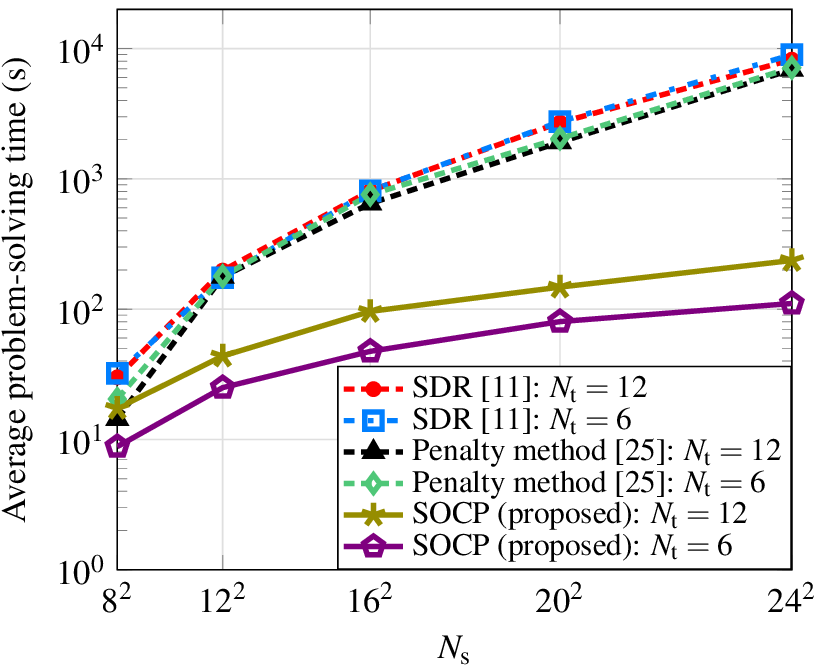}\vspace{-0.2cm}\caption{Average problem solving time versus $N_{\mathrm{s}}$ for $K=6$ and
$\Gamma=10$ dB.}
\label{fig:runTime_vs_Ns}
\end{figure}

\subsection{Application of the Proposed Method to Other Problems}

We remark that the proposed solution is dedicated to~\eqref{eq:Pn1},
but it has wider applicability. We now provide a brief discussion
on how the proposed solution can be applied to other IRS-related design
problems with slight modifications. The problem (10) in~\cite{21-TCOM-SISO-BC}
has exactly the same constraints as~\eqref{eq:SINRC-Orig} and~\eqref{eq:UMC-Orig},
and thus the proposed SOCP-based approach can directly be applied.
Also, the non-convex constraints of the PowerMin problem in~\cite{21-TCOM-SIMO-MAC}
are the same as~\eqref{eq:SINRC-Orig} and~\eqref{eq:UMC-Orig},
and thus Algorithm \ref{algo} is still directly applicable. Following
a similar line of arguments, one can easily see that the proposed
algorithm can be slightly modified to solve~\cite[(P1)]{21-WCNC-IoT}
and~\cite[(P1)]{21-TCOM-Renzo}, among the others.

\section{Numerical Results and Discussion}

We assume that the system operates at the center frequency of 2~GHz
with a total available bandwidth of 20~MHz. It is assumed that the
center of the BS uniform linear array and the center of the IRS uniform
planar array are located at $(0\ \mathrm{m},20\ \mathrm{m},10\ \mathrm{m})$
and $(30\ \mathrm{m},0\ \mathrm{m},5\ \mathrm{m})$, respectively.
The distance between the neighboring antennas at the BS, as well as
that between the neighboring elements at the IRS is equal to $\lambda/2$,
where $\lambda$ is the wavelength of the carrier waves. The single-antenna
users are assumed to be randomly distributed inside a circular disk
of radius $5\ \mathrm{m}$, centered at $(350\ \mathrm{m},10\ \mathrm{m},2\ \mathrm{m})$,
ensuring that the minimum distance between the users is equal to $2\lambda$.
We model the Rician distributed small-scale fading and distance-dependent
path loss for all of the wireless links according to~\cite[Sec.~VI]{21-Stefan-MIMOBC}.
The noise spectral density at each of the user node is assumed to
equal to -174dBm/Hz. Without loss of generality, it is assumed that
$\Gamma_{k}=\Gamma,\ \forall k\in\mathcal{K}$ and $\xi=0.001$. In
Figs.~\ref{fig:convergence}--\ref{fig:runTime_vs_Ns}, the averaging
is performed over 100 independent small-scale channel fading realizations,
and we consider 1000 independent vectors for Gaussian randomization
in each iteration for the SDR-based baseline algorithm. The simulations
are performed on a Linux PC with 7.5 GiB memory and Intel Core i5-7200U
CPU, using Python v3.9.7 and MOSEK Fusion API for Python Rel.-9.2.49~\cite{mosek}.\textcolor{black}{{}
It is noteworthy that when the value of the target SINR $\Gamma$
is small, the SINR constraints can be met easily. In such a scenario,
the impact of the SINR constraints on the optimization problem is
minimal, and therefore, different optimization schemes may achieve
similar performance. Therefore, we consider high/stricter SINR requirements
in our simulations.}

Fig.~\ref{fig:convergence} shows the convergence comparison of the
proposed SOCP- and SDR-based~\cite{19-TWC-Zhang-TPM} algorithms.
It is evident from the figure that Algorithm~\ref{algo} outperforms
the SDR-based algorithm in terms of required transmit power. More
interestingly, as the QoS constraints becomes more demanding (i.e.,
for large $\Gamma$), the performance gap between the two algorithms
increases. For large values of $\Gamma$, in fact, the likelihood
of obtaining a suitable Gaussian vector via randomization that satisfies
the QoS constraints decreases, resulting in the inferior performance
of SDR-based algorithm. In the remaining experiments, we stop Algorithm
\ref{algo} and the SDR-based method when either the relative change
in the objective is less than $10^{-5}$ or the number of iterations
is 20.

The impact of $N_{\mathrm{s}}$ on the average transmit power is shown
in Fig.~\ref{fig:power_vs_Ns}. \textcolor{black}{In this figure,
we also show the performance of a penalty-based AO method~\cite{22-secureISAC},
which can be modified to solve the PowerMin problem. As expected,
the average transmit power to maintain the desired QoS decreases as
$N_{\mathrm{s}}$ is increased. For large values of $N_{\mathrm{s}}$,
in fact, the IRS performs a highly-focused beamforming to achieve
the desired QoS even for small values of the transmit power. It is
evident from the figure that the SDR-based and penalty-based methods
perform quite similarly for small values of $\Gamma$. For large $\Gamma$,
however, the coupling among the optimization variables becomes stronger,
resulting in better performance for the penalty-based method compared
to the SDR-based method. Since all the design variables are updated
simultaneously, the SOCP-based method outperforms both the SDR- and
penalty-based methods relying on AO. }

In Fig.~\ref{fig:runTime_vs_Ns}, we show the average problem-solving-time
comparison between the proposed SOCP-based and SDR-based algorithms.
As expected, the required run-time increases with increasing values
of $N_{\mathrm{s}}$ for both the algorithms. Also, in line with the
complexity analysis in~Section~\ref{subsec:Complexity-Analysis},
the run-time for the SDR-based algorithm is much larger than that
for the proposed SOCP-based algorithm. \textcolor{black}{Finally,
we note that the per-iteration complexity of the method in~\cite{22-secureISAC}
is $\mathcal{O}(N_{\mathrm{s}}^{3})$, which is slightly less than
$\mathcal{O}(N_{\mathrm{s}}^{3.5})$ provided by our proposed method.
However, the method in~\cite{22-secureISAC} needs a significantly
larger number of iterations to converge (compared to the SOCP- and
SDR-based methods). Compared to the proposed SOCP-based method, thus,
it takes a much larger overall run time to return a solution, as shown
in Fig. \ref{fig:runTime_vs_Ns}.}

\section{Conclusion}

We have presented the joint transmit beamforming and IRS phase-shift
design to minimize the transmit power in an IRS-assisted MU-MIMO d\textcolor{black}{ownlink
communication system with QoS constraints. Specifically, we proposed
an SCA-assisted SOCP framework for solving the PowerMin problem where
all of the optimization variables are updated in each iteration, simultaneously,
in contrast to the conventional SDR- and penalty-based algorithms
in which only one set of optimization variables is updated in each
iteration with the others being fixed. The proposed algorithm was
shown to outperform the state-of-the-art benchmark schemes in terms
of both the required transmit power and the computational complexity/average
problem-solving time, and is promising to be applied to solve other
similar design problems in IRS-aided communications.}

\bibliographystyle{IEEEtran}
\bibliography{Kumar_WCL2022-2013}

\end{document}